\documentclass[12pt]{article}
\bigskip

\usepackage{amsfonts,amsmath,amssymb,amsbsy,amsthm}
\usepackage[bf]{caption}
\usepackage{cite}
\usepackage{color}
\usepackage{psfrag,graphicx}
\usepackage{shadow}
\usepackage{subfigure}
\usepackage{textcomp}
\usepackage{wrapfig}
\usepackage{upgreek}
\usepackage{rotating}

\DeclareSymbolFont{UPM}{U}{eur}{m}{n}
\DeclareMathSymbol{\upartial}{0}{UPM}{"40}

\newcommand{\fetlut}[1]{\mbox{\boldmath{\mbox{$#1$}}}} 

\newcommand{\dirac}[1]{\updelta\!\left({#1}\right)}

\newcommand{\vfi}{\varphi}

\newcommand{\ds}{\displaystyle}

\renewcommand{\j}{\text{j}}

\newcommand{\jw}{\j\omega}
\newcommand{\e}{\text{e}}

\newcommand{\br}[1]{\left({#1}\right)}

\newcommand{\brn}[1]{\!\left({#1}\right)} 

\newcommand{\brh}[1]{\left[{#1}\right]}
\newcommand{\brc}[1]{\left\{{#1}\right\}}
\newcommand{\brb}[1]{\left|{#1}\right|}
\newcommand{\bri}[1]{\left\langle{#1}\right\rangle}

\newcommand{\lap}{\nabla^2}

\newcommand{\rot}{\nabla\times}

\newcommand{\grad}{\nabla}

\renewcommand{\div}{\nabla\cdot}

\newcommand{\pden}[2]{\ds\frac{\upartial #1}{\upartial #2}}

\newcommand{\pdx}[1]{\ds\frac{\upartial #1}{\upartial x}}
\newcommand{\pdy}[1]{\ds\frac{\upartial #1}{\upartial y}}

\newcommand{\pdro}[1]{\ds\frac{\upartial #1}{\upartial \rho}}

\newcommand{\pdvf}[1]{\ds\frac{\upartial #1}{\upartial \varphi}}

\newcommand{\dt}[2]{\ds\frac{\dd #1}{\dd #2}}

\newcommand{\vc}[1]{\fetlut{#1}}   

\newcommand{\uv}[1]{\hat{\vc{#1}}} 
\newcommand{\un}{\uv{n}}

\newcommand{\ux}{\uv{x}}
\newcommand{\uy}{\uv{y}}
\newcommand{\uz}{\uv{z}}

\newcommand{\uro}{\uv{\rho}}

\newcommand{\uvf}{\uv{\varphi}}

\newcommand{\Avv}{\vc{A}}
\newcommand{\Bvv}{\vc{B}}
\newcommand{\Cvv}{\vc{C}}

\newcommand{\Fvv}{\vc{F}}

\newcommand{\Ivv}{\vc{I}}
\newcommand{\Jvv}{\vc{J}}
\newcommand{\Kvv}{\vc{K}}

\newcommand{\bvv}{\vc{b}}
\newcommand{\cvv}{\vc{c}}

\newcommand{\kvv}{\vc{k}}

\newcommand{\rvv}{\vc{r}}

\DeclareMathAlphabet{\mathsfbfsl}{T1}{cmss}{bx}{sl}

\newcommand{\dd}{\mbox{d}}

\newcommand{\dS}{\dd s}

\newcommand{\dl}{\dd l}

\begin{document}

\title{{\Large \bf Explicit reconstruction of line-currents and their positions in a two-dimensional parallel conductor structure}}

\author{Martin Norgren}
\maketitle

\begin{center}
Dep. of Electromagnetic Engineering\\ Royal Institute of
Technology\\ SE-100 44 Stockholm, Sweden \\[6mm]
Corresponding author: Martin Norgren\\
Email: martin.norgren@ee.kth.se\\ Tel: +46 8 7907410; Fax: +46 8
205268\\[10mm]
\end{center}

\newpage

\begin{abstract}
 The magnetic inverse source problem of reconstructing the positions
 and currents of very long parallel conductors is considered in a two-dimensional situation, with applications to power line measurements. The input data is the magnetic field on a contour surrounding the conductors to be reconstructed. Using a scalar-vector Green identity, an explicit reconstruction algorithm is derived. The numerical implementation of the algorithm is described and simulation results are presented, demonstrating the influences from numerical errors and uncertainties in measurement data. The algorithm can handle an arbitrary number of conductors, but stability problems associated with the illposedness accelerate with increasing number of conductors. Mathematically, the Green identity approach removes the influence of external disturbances and thus have potential usefulness in current reconstruction for determining optimal sensor positions and how to process measurement data. \medskip

\end{abstract}


\section{Introduction}

In electric power engineering, magnetic inverse source problems appear in contact-free measurements of currents in power lines \cite{EMANUEL_ETAL:1983,CYGANSKI_ETAL:1984}, bus bars \cite{DIRIENZO_ETAL:2001,ZHANG_DIRIENZO:2009,DIRIENZO_ZHANG:2010} and cables \cite{BELLINA_ETAL:2001,BRUZZONE_ETAL:2002,FORMISANO_ETAL:2005}. The magnetic field is measured by a set of sensors located in the vicinity of the conductors, and the currents are determined by inversion of the direct map from the currents to the magnetic field, using a suitable model of the problem.  Due to the illposedness of such inverse problems, the results become highly sensitive to errors in the measured magnetic field and to deficiencies in the models. Hence, much effort has been spent on reducing the impact of disturbances from exterior magnetic fields \cite{DIRIENZO_ETAL:2001,DIRIENZO_ZHANG:2010}, and on finding optimal sensor positions \cite{BRUZZONE_ETAL:2002,ZHANG_DIRIENZO:2009}. In electric power engineering, the focus has been on reconstructing the currents in individual conductors. Owing to the fact that in many situations the conductor positions are known or determinable by other means than magnetic field measurements, the simultaneous reconstruction of positions and currents has devoted less interest. An exception is the early work in \cite{CYGANSKI_ETAL:1984}, where also the conductor positions are reconstructed, from experimental data.

Another application of magnetic inverse source problems is in biomagnetics. For example, in magnetoencephalography (MEG) epileptic activity is detected as current sources within the human brain \cite{HAMALAINEN_ETAL:1993}. With MEG being a tool providing input data to radiotherapy or surgical treatment, usually the localization of the current source is more important than the detailed determination of the current.
In e.g. MEG, the localization problem can sometimes be solved explicitly by means of methods based on using harmonic functions in combination with Green identities \cite{HE_ROMANOV:1998,ELBADIA_HADUONG:2000,POPOV:2002}. Such explicit methods typically require measurement data on closed surfaces that completely surround the region containing the sources. In MEG, magnetic field data can be obtained over a quite large solid angle, which together with data-continuation facilitates explicit identification methods \cite{POPOV:2002}. However, an obstacle is that the current sources are submerged into a conducting medium while the magnetic field is measured in another, non-conducting, region that typically is air. Hence, for quasistatic fields, measurement data cannot be continued into the source region, in a stable manner. One way around that problem is to use a simplistic model (e.g. model the brain as a sphere) that allows the configuration to be replaced by an equivalent source in free space    \cite{He_Norgren:2000}. However, such an approach may introduce large modelling errors.

In many electric power applications, the conductors are located in free space, i.e. the same region in which the surrounding magnetic field is measured. Thus, one will not encounter the difficulties with data continuation that appears in MEG problems. Hence, it is of interest to investigate whether the explicit localization methods, considered previously in MEG, can be used in power engineering for reconstructing currents and localizations of conductors, in overhead power lines, bus bars etc.

In this paper, we consider the two-dimensional approximation of the magnetic inverse source problem of reconstructing the positions and currents of very long parallel conductor structures. In section \ref{sekt:Problem formulation and preliminaries}, we formulate the problem and present the theory behind the explicit method. The reconstruction algorithm is described in Section \ref{sekt:Rekalg}. The numerical implementation and simulations results are presented in Section \ref{sekt:Reconstruction results}, and Section \ref{sekt:Discussion and conclusions} contains the conclusions.

\section{Problem formulation and preliminaries}

\label{sekt:Problem formulation and preliminaries}

\begin{figure}[h]
\centering \psfrag{x}{$x$} \psfrag{y}{$y$} \psfrag{z}{$z$}
\psfrag{i}{$\Jvv$} \psfrag{e}{$\Jvv$}
\psfrag{n}{$\un$} \psfrag{t}{$\uv{t}$}\psfrag{S}{$\mathcal{S}$} \psfrag{C}{$\mathcal{C}$}
\subfigure[General longitudinal current density in the reconstruction region.]
{
\label{Fig:1a}
\includegraphics[scale=0.9]{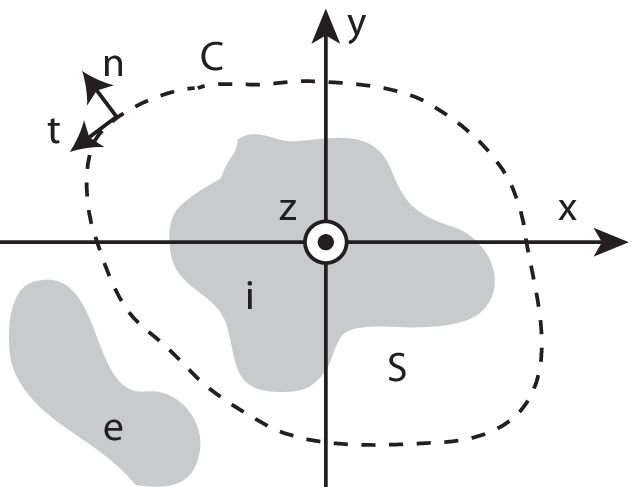}
}
\hspace{10mm}
\subfigure[Line-current model of the current density in the reconstruction region.]
{
\label{Fig:1b}
\psfrag{x}{$x$} \psfrag{y}{$y$} \psfrag{z}{$z$}
\psfrag{J}{$\Jvv$} \psfrag{e}{$\Jvv_\text{e}$}
\psfrag{n}{$\un$} \psfrag{S}{$\mathcal{S}$} \psfrag{C}{$\mathcal{C}$}
\psfrag{1}{$I_1$} \psfrag{2}{$I_2$} \psfrag{3}{$I_N$}
\includegraphics[scale=0.9]{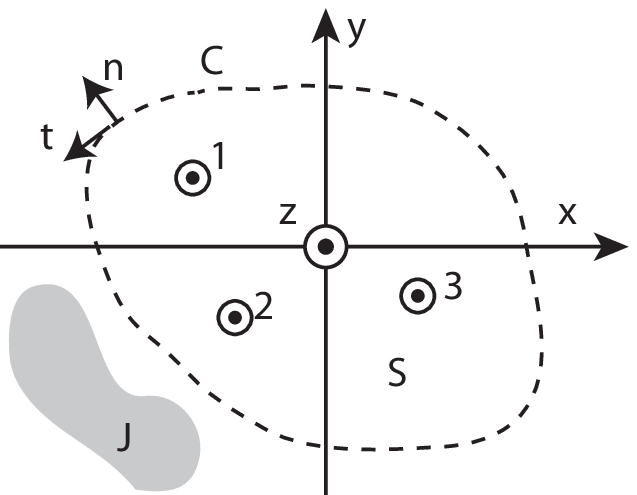}
}
\caption{The problem geometry. The inverse problem is to reconstruct the current density inside the contour $\mathcal{C}$ using magnetic field data on $\mathcal{C}$. $\mathcal{S}$ is the surface bounded by $\mathcal{C}$, $\un$ and $\uv{t}$ are the unit normal and unit tangent, respectively, on $\mathcal{C}$.  }\label{Fig:1}
\end{figure}

We consider a two-dimensional approximation of a parallel conductor structure. The currents distributed over the cross section of the structure are modelled as a two-dimensional longitudinal current density
\begin{align}
\Jvv\brn{\rvv}=J\brn{x,y}\uz
\label{eq:Jdef}
\end{align}
where $\rvv=x\ux+y\uy$. $\ux, \uy$ and $\uz$ denote the unit vectors in the cartesian coordinates; see Figure \ref{Fig:1}.
The current density $\Jvv$ may consist of several disjoint parts and the inverse problem is to reconstruct the part internal to a {\em measurement contour} $\mathcal{C}$ by using magnetic field data on $\mathcal{C}$.

In the {\em reconstruction region} $\mathcal{S}$, the surface bounded by $\mathcal{C}$, we have no access to magnetic field data. It is assumed that $\mathcal{C}$ passes between disjoint parts of $\Jvv$, which means that on and in the neighborhood of $\mathcal{C}$ we have $\Jvv=\vc{0}$.
In practice, the current density  exterior to $\mathcal{C}$ represent nearby parallel conductors and/or the influence from a parallel ground.

The preliminary analysis is carried out assuming that $\Jvv$ has a general distribution in the reconstruction region; see Figure \ref{Fig:1a}. The reconstruction algorithm, described in Section \ref{sekt:Rekalg}, is restricted to the case when $\Jvv$  is a finite number of line-currents; see Figure \ref{Fig:1b}.
Assuming slowly varying currents, we neglect propagation effects and apply static analysis methods. Hence, the magnetic field $\Bvv$ fulfils
\begin{align}
\div\Bvv\brn{\rvv} & = 0 \label{eq:divB}\\
\rot\Bvv\brn{\rvv} & = \mu_0\Jvv\brn{\rvv} \label{eq:rotB}
\end{align}
where $\grad=\ux\upartial/\upartial x + \uy\upartial/\upartial y$. From \eqref{eq:Jdef}, \eqref{eq:divB} and \eqref{eq:rotB} it follows that $\Jvv$ generates a transversal magnetic field: $\Bvv = B_x\ux + B_y\uy$. If we measure amplitudes and phases of time-harmonic fields, instead of instantaneous values, $\Jvv$ and $\Bvv$ represent complex phasors, with the time-dependence $\exp\brn{\jw t}$ assumed and suppressed.

\subsection{Integral equation for $\Jvv$}

Consider the following Green identity for a scalar and a vector function \cite{STAFFAN:1991}:
\begin{align}
\int_\mathcal{S}  & \brh{G\lap\Bvv  - \Bvv\lap G}\dS \nonumber \\
& \qquad = \oint_\mathcal{C}\brh{G\br{\div\Bvv}\un-\br{\un\cdot\Bvv}\grad G + G \br{\rot\Bvv}\times\un - \br{\un\times\Bvv}\times\grad G}\dl \label{eq:Greensksats1}
\end{align}
where the scalar function $G$ is a  function to be specified and $\un$ is the from $\mathcal{S}$ outwardly directed unit normal on $\mathcal{C}$; see Figure \ref{Fig:1}. \eqref{eq:divB} holds generally, and with no current density on $\mathcal{C}$ \eqref{eq:rotB} implies $\rot\Bvv=\vc{0}$ on $\mathcal{C}$. Thus, \eqref{eq:Greensksats1} reduces to
\begin{align}
\int_\mathcal{S}   \brh{G\lap\Bvv  - \Bvv\lap G}\dS
  = - \oint_\mathcal{C}\brh{\br{\un\cdot\Bvv}\grad G + \br{\un\times\Bvv}\times\grad G}\dl \label{eq:Greensksats2}
\end{align}
which means that on $\mathcal{C}$ we do not need to measure derivatives of the magnetic field. However, over $\mathcal{S}$ the magnetic field is not accessible for measurements, and to handle that in \eqref{eq:Greensksats2} we require the function $G$ to be harmonic over $\mathcal{S}$, i.e.
\begin{align}
\lap G = 0
\label{eq:lapG}
\end{align}
Hence, \eqref{eq:Greensksats2} simplifies to
\begin{align}
\int_\mathcal{S}   G\lap\Bvv \dS
  = - \oint_\mathcal{C}\brh{\br{\un\cdot\Bvv}\grad G + \br{\un\times\Bvv}\times\grad G}\dl \label{eq:Greensksats3}
\end{align}
From \eqref{eq:divB} and \eqref{eq:rotB} it follows that the Poisson equation for $\Bvv$ becomes
\begin{align}
\lap\Bvv = \grad\br{\div\Bvv}-\rot\br{\rot\Bvv} = - \mu_0 \rot\Jvv \label{eq:lapB}
\end{align}
which, by integration by parts over $\mathcal{S}$ utilizing that $\Jvv=\vc{0}$ on $\mathcal{C}$, yields that the left hand side of \eqref{eq:Greensksats3}
\begin{align}
\int_\mathcal{S}   G\lap\Bvv \dS = -\mu_0\int_\mathcal{S}   G\rot\Jvv \dS = -\mu_0\int_\mathcal{S}   \Jvv\times\grad G \dS
\label{eq:Greensksats4}
\end{align}
With $\Jvv\times\grad G  = J \uz\times\grad G$ and $-\uz\times\brn{\uz\times\grad G}=\grad G$, we obtain using \eqref{eq:Greensksats3} and \eqref{eq:Greensksats4} that
\begin{align}
\int_\mathcal{S}   J \grad G \dS
  = -\frac{1}{\mu_0} \uz\times\oint_\mathcal{C}\brh{\br{\un\cdot\Bvv}\grad G + \br{\un\times\Bvv}\times\grad G}\dl \label{eq:Greensksats6}
\end{align}
Considering the right hand side as known, \eqref{eq:Greensksats6} is an integral equation for the current density $J\brn{\rvv}$, where the gradient of the harmonic function $G\brn{\rvv}$ is the kernel.

\subsection{The choice of the harmonic function $G$}

Aiming at an explicit reconstruction algorithm, the harmonic function $G\brn{\rvv}$ in \eqref{eq:Greensksats6} must be reasonably simple. Similar to in \cite{HE_ROMANOV:1998,He_Norgren:2000}, $G$ is defined as
\begin{align}
G\brn{\rvv} & = G\brn{w\brn{\rvv}}, \label{eq:Gw}\\
w & = \kvv\cdot\rvv \label{eq:wekv}
\end{align}
where the complex vector $\kvv=k_x\ux + k_y\uy$ fulfils $\kvv\cdot\kvv=0$.
Denoting $\dd{G}/\dd{w}=f\brn{w}$, we obtain
\begin{align}
\grad G\brn{\rvv} & = f\brn{w} \grad w\brn{\rvv} = f\brn{w} \kvv \label{eq:gradG}\\
 \lap G\brn{\rvv} & = \div\grad G\brn{\rvv} = \kvv\cdot\grad f\brn{w\brn{\rvv}} = \dt{f}{w}\kvv\cdot\grad w\brn{\rvv} = \dt{f}{w}\kvv\cdot\kvv = 0\label{eq:lapGk}
\end{align}
Hence, for \eqref{eq:lapG} to hold, the vector $\kvv$ must be complex. For three dimensions, the general form of $\kvv$ is given in \cite{HE_ROMANOV:1998}. For two dimensions, we conclude that the general form of $\kvv$ is
\begin{align}
\kvv = \alpha\br{\ux + \j s \uy}\label{eq:kekv}
\end{align}
where $\alpha\in\mathbb{C}$ is an arbitrary constant and $s=\pm 1$. Using \eqref{eq:gradG} in \eqref{eq:Greensksats6}, it follows that
\begin{align}
\kvv\int_\mathcal{S}   J f \dS
  = -\frac{1}{\mu_0} \uz\times\oint_\mathcal{C}\brh{\br{\un\cdot\Bvv}\kvv + \br{\un\times\Bvv}\times\kvv}f\dl \label{eq:Greensksats7}
\end{align}
Since the two components of \eqref{eq:Greensksats7} are in effect two identical equations, we dot-multiply \eqref{eq:Greensksats7} with $\kvv^\ast$ ($\ast$ denotes the complex conjugate) which after some algebra yields
\begin{align}
\int_\mathcal{S}   J f \dS
  = \frac{1}{\mu_0} \oint_\mathcal{C}\brh{\j s \un\cdot\Bvv + \uv{t}\cdot\Bvv}f\dl \label{eq:Greensksats8}
\end{align}
Since we are working in two dimensions, it follows from \eqref{eq:wekv} and \eqref{eq:kekv} that in \eqref{eq:Gw}
\begin{align}
w = \alpha\br{x+\j s y}
\end{align}
which is an analytic function of either of the complex variables $x+\j y$ or $x-\j y$. Hence, \eqref{eq:lapG} becomes fulfilled if $G\brn{w}$ is an analytic function for all $w$ corresponding to $\rvv\in\mathcal{S}$. Consequently, $f\brn{w}=\dd {G}/\dd{w}$ is also an analytic function.

Note that with $f=$ constant, the property $\oint_\mathcal{C}\un\cdot\Bvv\dl = \int_\mathcal{S}\div\Bvv\dS = 0$ yields that \eqref{eq:Greensksats8} reduces to the well-known circulation law $\oint_\mathcal{C}\Bvv\cdot\uv{t}\dl = \mu_0 \int_\mathcal{S}\Jvv\cdot\uz\dS $. Hence, in this two-dimensional situation, \eqref{eq:Greensksats8} can be considered as a generalized circulation law, through which more information about the enclosed current density can be obtained by altering the harmonic function $f$.

\section{Algorithm for reconstructing line-currents}

\label{sekt:Rekalg}

Due to the existence of so-called silent or non-radiating sources, which produce zero fields outside their region of support, generally formulated inverse source problems typically exhibit non-unique solutions; a property utilized practically in so-called magnetic signature reduction \cite{NORGREN_HE_DEGAUSS:2000}.
In Appendix \ref{app:tyst}, we give examples of silent current densities that cannot be determined using the integral equation \eqref{eq:Greensksats6}. One way to achieve uniqueness is to shrink the space of possible solutions.

From now on, we restrict the interior current density $J\brn{\rvv}$ to be a finite number $N$ of line-currents:
\begin{align}
J\brn{\rvv} = \sum_{n=1}^N I_n \dirac{\rvv-\rvv_n} \label{eq:Imodell}
\end{align}
where $\brc{I_n}_{n=1}^N$ are the currents and $\brc{\rvv_n}_{n=1}^N$ are the conductor positions; see Figure \ref{Fig:1b} ($\dirac{}$ is the Dirac-delta function). The inverse problem thus becomes to reconstruct the currents and the conductor positions. In this study, we assume that $N$ is known; for the determination of $N$, see \cite{ELBADIA_HADUONG:2000}.

With $\Jvv=J\uz$, we obtain from \eqref{eq:lapB} and \eqref{eq:Imodell} that in the reconstruction region $\mathcal{S}$
\begin{align}
\lap B_x & = - \mu_0 \sum_{n=1}^N I_n \pdy{}\dirac{\rvv-\rvv_n}\\
\lap B_y & =   \mu_0 \sum_{n=1}^N I_n \pdx{}\dirac{\rvv-\rvv_n}
\end{align}
Hence, mathematically, the interior source of the field components is a finite number of dipolar sources, and in \cite{ELBADIA_HADUONG:2000} it has been shown that this kind of inverse problem has a unique solution.

The reconstruction algorithm has originally been derived in \cite{ELBADIA_HADUONG:2000}. Here, we give an alternative derivation, adapted to the present context. With $f\brn{w}$ being an analytic function it follows that $\brh{f\brn{w}}^m, m=0, 1, 2, \ldots$ are analytic functions as well. Let $f_n^m = \brh{f\brn{w_n}}^m$, where $w_n=\kvv\cdot\rvv_n$. Hence, we obtain from \eqref{eq:Greensksats8} and \eqref{eq:Imodell} that
\begin{align}
\sum_{n=1}^N I_n f_n^m
  = \frac{1}{\mu_0} \oint_\mathcal{C}\brh{\j s \un\cdot\Bvv + \uv{t}\cdot\Bvv}f^m \dl = b_m,\qquad m=0,1,2,\ldots \label{eq:Greensksats9}
\end{align}
where $b_m$ are constants determined from the magnetic field measurement on $\mathcal{C}$ and the choice of $f\brn{w\brn{\rvv}}$. In \eqref{eq:Greensksats9}, we have $2N$ unknowns in $\brc{I_n, f_n}_{n=1}^N$. If $\brc{f_n}_{n=1}^N$ have been determined, the currents can be obtained from a linear equation system generated by $N$ different values on $m$. For example, using $m=M, M+1,\ldots, M+N-1$ (where $M\geq 0$), we introduce the following matrix and vectors:
\begin{align}
\Fvv_M = \begin{bmatrix} f_1^M & \cdots & f_N^M \\
\vdots & \ddots & \vdots \\
f_1^{N-1+M} & \cdots & f_N^{N-1+M}
\end{bmatrix},\qquad \Ivv = \begin{bmatrix} I_1 \\
\vdots  \\
I_N
\end{bmatrix},\qquad \bvv_M = \begin{bmatrix} b_M \\
\vdots  \\
b_{N-1+M}
\end{bmatrix}
\end{align}
Hence, if $\Fvv_M$ is known the current vector $\Ivv$ is determined from the equation
\begin{align}
\Fvv_M \vc{I} = \bvv_M \label{eq:Iekv}
\end{align}
The complex numbers $\brc{f_n}_{n=1}^N$ are the roots of a polynomial equation:
\begin{align}
f^N + c_{N-1}f^{N-1}+ \ldots + c_1 f + c_0 = 0 \label{eq:fnekv}
\end{align}
Introducing the vector $\cvv$ containing the polynomial coefficients:
\begin{align}
\cvv = \begin{bmatrix} c_0 \\
\vdots  \\
c_{N-1}
\end{bmatrix}\label{eq:cekv}
\end{align}
we obtain using \eqref{eq:fnekv} that
\begin{align}
\cvv^\text{T} \Fvv_M = - \begin{bmatrix} f_1^{N+M} \cdots f_N^{N+M}\end{bmatrix}
\label{eq:cTFM}
\end{align}
(T denotes matrix transpose). Hence, using \eqref{eq:Iekv} and \eqref{eq:cTFM}, the scalar
\begin{align}
\bvv_M^\text{T} \cvv = \cvv^\text{T} \bvv_M = \cvv^\text{T} \Fvv_M \Ivv = - \begin{bmatrix} f_1^{N+M} \cdots f_N^{N+M}\end{bmatrix} \Ivv = - b_{M+N}
\end{align}
Now, using $N$ different values of $M$ in the relation $\bvv_M^\text{T} \cvv = - b_{M+N}$, we obtain a linear system from which the coefficient vector $\cvv$ can be determined. For example, using $M=L,\ldots, L+N-1$ (where $L\geq 0$), we construct the matrix
\begin{align}
\Cvv_L = \begin{bmatrix} \bvv_L^\text{T}\\ \vdots \\ \bvv_{L+N-1}^\text{T} \end{bmatrix}
\end{align}
in which way the coefficient vector $\cvv$ follows from the equation
\begin{align}
\Cvv_L \cvv = - \bvv_{L+N}
\end{align}
With $\cvv$ determined, $\brc{f_n}_{n=1}^N$ follows from \eqref{eq:fnekv}, whereafter $\brc{I_n}_{n=1}^N$ follows from \eqref{eq:Iekv}.

\section{Reconstruction results}

\label{sekt:Reconstruction results}

\subsection{Numerical approximation of the measurement integral and reconstructions from clean data}
\begin{figure}[h]
\centering \psfrag{x}{$x$} \psfrag{y}{$y$} \psfrag{z}{$z$}
\psfrag{J}{$\Jvv$} \psfrag{e}{$\Jvv_\text{e}$}
\psfrag{n}{$\un$} \psfrag{S}{$\mathcal{S}$} \psfrag{C}{$\mathcal{C}$}
\psfrag{1}{$I_1$} \psfrag{2}{$I_2$} \psfrag{3}{$I_N$}
\subfigure[Measurement contour and conductor locations used in the numerical examples.]
{
\label{Fig:3a}
\includegraphics[scale=0.9]{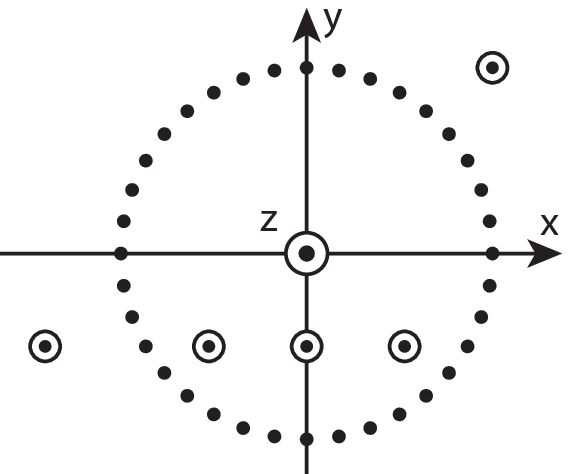}
}
\hspace{10mm}
\subfigure[A straight segment $\Delta\mathcal{C}$ of $\mathcal{C}$ between two measurement points.]
{
\label{Fig:3b}
\psfrag{1}{$\rvv_1$} \psfrag{2}{$\rvv_2$} \psfrag{t}{$\uv{t}$} \psfrag{c}{$\Delta\mathcal{C}$}
\includegraphics[scale=1.2]{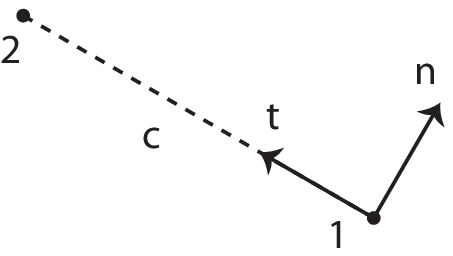}
}
\caption{ }\label{Fig:3}
\end{figure}

The input to the reconstruction algorithm is the $b_m$-coefficients, obtained by integration over the measurement contour $\mathcal{C}$  (see \eqref{eq:Greensksats8}):
\begin{align}
b_m = \frac{1}{\mu_0} \oint_\mathcal{C}\brh{\j s \un\cdot\Bvv + \uv{t}\cdot\Bvv}f^m \dl, \qquad m=0,1,2,\ldots  \label{eq:bmekv}
\end{align}
In a practical situation, magnetic field data is obtained only at a finite number of points, wherefore \eqref{eq:bmekv} must be evaluated approximately. Here, we choose to distribute the measurement points uniformly around a circle with the radius $R_\text{meas}$; see Figure \ref{Fig:3a}.

Since the integral in \eqref{eq:bmekv} is independent of the curve $\mathcal{C}$ as long as it circumscribes only the currents to be reconstructed, we chose  $\mathcal{C}$ as straight segments between the measurement points. On a segment $\Delta\mathcal{C}$ between points $\rvv_1$ and $\rvv_2$, see Figure \ref{Fig:3b}, we obtain
\begin{align}
\rvv\brn{t}  = \rvv_1\br{1-t} + \rvv_2 t, \qquad
\dl  = \brb{\rvv_2 - \rvv_1}\dd t
\end{align}
where the parameter $t\in\brh{0,1}$. At $\rvv_1$ and $\rvv_2$, the measured magnetic fields are denoted $\Bvv_1$ and $\Bvv_2$, respectively. On $\Delta\mathcal{C}$ the magnetic field is interpolated linearly:
\begin{align}
\Bvv\brn{t} & = \Bvv_1\br{1-t} + \Bvv_2 t, \qquad \rvv\brn{t}\in\Delta\mathcal{C}\label{eq:Binterp}
\end{align}
Utilizing that $\un=\uv{t}\times\uz$, the contribution to \eqref{eq:bmekv} from $\Delta\mathcal{C}$ becomes
\begin{align}
\Delta b_m = \frac{\brb{\rvv_2 - \rvv_1}}{\mu_0}\int_0^1\brh{B_x\brn{t}\br{t_x+\j s t_y} + B_y\brn{t} \br{t_y - \j s t_x}}\brh{f\brn{\rvv\brn{t}}}^m \dd t \label{eq:Dbmekv}
\end{align}
Even though many choices of $f\brn{t}$ admit explicit evaluation of the integral in \eqref{eq:Dbmekv}, we find it from the versatility point of view better to use a numerical quadrature routine; the numerical accuracy is anyhow superior to the approximation in \eqref{eq:Binterp}.

The analytic function $f\brn{w}=f\brn{x+\j y}$, we chose as
\begin{align}
f\brn{x,y} = \exp\brn{\frac{\j x -y}{R_\text{meas}}}
\end{align}
Hence, once the evaluations $f_n$ at the conductor positions have been found, the conductor coordinates are obtained as
\begin{align}
x_n & = R_\text{meas}\text{Im}\brc{\ln f_n}\\
y_n & = - R_\text{meas}\text{Re}\brc{\ln f_n}
\end{align}
Note that, in addition to being analytic, the function $f\brn{w}$ must be invertible in the reconstruction region $\mathcal{S}$.

As the test case, we consider three internal conductors, whose positions and currents are to be reconstructed, and two external conductors that act as disturbance sources. The coordinates and phasor currents of the conductors are given in Table \ref{tab:leddata} (conductor positions also illustrated in Figure \ref{Fig:3a}).
\begin{table}[h]
\caption{Locations and currents for the conductors used in the numerical examples.}
\centering
{
\begin{tabular}{|l||p{12mm}|p{12mm}|p{12mm}||p{12mm}|p{12mm}|}
\hline
Conductor data & \multicolumn{3}{p{36mm}||}{Inside $\mathcal{C}$} & \multicolumn{2}{p{24mm}|}{Outside $\mathcal{C}$}\\
\hline
Coordinate $x/R_\text{meas}$ & -0.5 & 0 & 0.5 & -1.5 & 1 \\
\hline
Coordinate $y/R_\text{meas}$ & -0.5 & -0.5 & -0.5 & -0.5 & 1 \\
\hline
Current $I/$A & -\j & 2 & 1 & -1 & 2\j \\
\hline
\end{tabular}}\label{tab:leddata}
\end{table}

For the generation of artificial magnetic field data, we use the formula
\begin{align}
\Bvv\brn{\rvv} = \frac{\mu_0}{2\pi}\sum_{n=1}^{N+N_\text{ext}} \frac{I_n\uz\times\br{\rvv-\rvv_n}}{\brb{\rvv-\rvv_n}^2}
\label{eq:Bdirekt}
\end{align}
To check the accuracy of the linear interpolation, we take the input data from Table \ref{tab:leddata} and use \eqref{eq:Bdirekt} in \eqref{eq:Binterp} and \eqref{eq:Dbmekv}, and compare with the exact results, obtained by using the input data in the left hand side of \eqref{eq:Greensksats9}. For a smooth magnetic field, the dominant error in the linear interpolation \eqref{eq:Binterp} is quadratic in the segment length, wherefore the rate of convergence of the integral for $b_m$ is expected to be proportional to the inverse square of the number of measurement points; this was also supported by observations. Hence, we use Richardson extrapolation to improve the numerical accuracy. We sample the circular measurement curve at every $10^\circ$, as depicted in Figure \ref{Fig:3a}, and let $b_m^\text{all}$ denote the results when using all 36 measurement points. Then we use data from 18 points at even and odd multiples of $10^\circ$ and denote the corresponding results $b_m^\text{even}$ and $b_m^\text{odd}$, respectively. The average of the extrapolations from the even and odd sets becomes
\begin{align}
b_m^\text{extrapol} = \frac{8 b_m^\text{all} - b_m^\text{even} - b_m^\text{odd}}{6}\label{eq:extrapol}
\end{align}
In Table \ref{tab:interpol}, we compare the results, when using $m\in\brc{1,2,3,4,5,6}$; the necessary and sufficient number of $m$-values for reconstructing three line-currents. We conclude that the Richardson extrapolation improves significantly the numerical accuracy.

\begin{table}[h]
\caption{Results, in the unit A, of extrapolations of the coefficients $b_m$, evaluated by \eqref{eq:Dbmekv}. $b_m^\text{even}$ are from measurement points at $\brc{0^\circ, 20^\circ,\ldots 340^\circ}$, on the circular measurement curve $\mathcal{C}$. $b_m^\text{odd}$ are from measurement points at $\brc{10^\circ, 30^\circ,\ldots 350^\circ}$. $b_m^\text{all}$ are from all measurement points. $b_m^\text{extrapol}$ are the extrapolations using \eqref{eq:extrapol}. $b_m^\text{exact}$ are the exact results, calculated directly from the conductor currents. $m$ is the power of the function $f$.}
{\begin{tabular}{|l|r|r|r|r||r|}
\hline
$m$ & $b_m^\text{even}$ & $b_m^\text{odd}$ & $b_m^\text{all}$ & $b_m^\text{extrapol}$ &  $b_m^\text{exact}$ \\
\hline
1 &   1.027\! - \!2.132j\! &  1.027\! - \!2.133j\! &  1.052\! - \!2.211j\! &  1.060\! - \!2.237j\! &  1.060\! - \!2.237j \\ \hline
2 &   1.494\! - \!3.575j\! &  1.494\! - \!3.573j\! &  1.630\! - \!3.711j\! &  1.675\! - \!3.757j\! &  1.681\! - \!3.756j \\ \hline
3 &   3.302\! - \!4.793j\! &  3.312\! - \!4.779j\! &  3.937\! - \!4.797j\! &  4.147\! - \!4.801j\! &  4.176\! - \!4.787j \\ \hline
4 &   8.215\! - \!4.648j\! &  8.248\! - \!4.612j\! & 10.331\! - \!3.934j\! & 11.031\! - \!3.702j\! & 11.134\! - \!3.644j \\ \hline
5 &  19.169\! - \!1.171j\! & 19.238\! - \!1.112j\! & 24.727\! + \!1.457j\! & 26.568\! + \!2.324j\! & 26.834\! + \!2.469j \\ \hline
6 & \!\! 40.514\!\! + \!\!8.492j\! & \!\! 40.608\!\! + \!\!8.565j\! & \!\! 52.670\!\! + \!\!14.760j\! & \!\! 56.706\!\! + \!\!16.837j\! & \!\! 57.221\!\! + \!\!17.050j\! \\ \hline
\end{tabular}}
\label{tab:interpol}
\end{table}

To evaluate the stability of the reconstruction algorithm against the numerical approximations, we first test the algorithm using clean measurement data. Referring to Section \ref{sekt:Rekalg}, we use the values $N=3, M=L=1$. We chose three different values of the number, $N_\text{meas}$, of measurement points. For the internal conductors, the results for the positions are presented in Table \ref{tab:renadataxy} and the results for the currents are presented in Table \ref{tab:renadataI}.  We observe that each doubling of the number of measurement points reduces effectively the errors in the reconstructions.
\begin{table}[h]
\caption{Reconstruction of conductor positions when using clean magnetic field data, from 72, 36 and 18 measurement points, respectively. The results are presented as displacements from the true values, in units of $R_\text{m}$, the radius of the measurement curve.}
\centering
{\begin{tabular}{l|l|l|l||l|l|l|}
\cline{2-7}

 & \multicolumn{3}{|c||}{$x_\text{true}/R_\text{meas}$} & \multicolumn{3}{|c|}{$y_\text{true}/R_\text{meas}$} \\ \cline{2-7}

    & -0.5 & 0 & 0.5 & -0.5 & -0.5 & -0.5  \\ \cline{2-7} \hline\hline
\multicolumn{1}{|c||}{$N_\text{meas} $} & \multicolumn{3}{|c||}{$\Delta x$ in \% of $R_\text{meas}$} & \multicolumn{3}{|c|}{$\Delta y$ in \% of $R_\text{meas}$}  \\ \hline
\multicolumn{1}{|c||}{72}  & 0.01 & 0.04 & -0.02 & 0.05 & -0.03 & 0.04  \\ \hline
\multicolumn{1}{|c||}{36}  & 0.16 & 0.60 & -0.24 & 0.65 & -0.40 & 0.64  \\ \hline
\multicolumn{1}{|c||}{18}  & 1.79 & 4.25 & -1.30 & 6.08 & -3.52 & 5.91  \\ \hline
\end{tabular}}
\label{tab:renadataxy}
\end{table}

\begin{table}[h]
\caption{Reconstruction results for conductor currents when using clean magnetic field data, from 72, 36 and 18 measurement points, respectively.}
\centering
{\begin{tabular}{|l|l|l|l|}
\hline
Currents/A   & $I_1$/A  & $I_2$/A  & $I_3$/A  \\ \hline
True values  &   0-j &    2+0j    &       -1+0j  \\ \hline
$N_\text{meas}=72$  &  0.0019 - 1.0008j & 1.9993 - 0.0009j & -1.0012 + 0.0017j \\ \hline
$N_\text{meas}=36$  &  0.0274 - 1.0091j & 1.9874 - 0.0149j & -1.0145 + 0.0243j \\ \hline
$N_\text{meas}=18$  &  0.2489 - 1.0419j & 1.7706 - 0.1944j & -1.0187 + 0.2384j
 \\ \hline
\end{tabular}}
\label{tab:renadataI}
\end{table}

\subsection{Reconstructions from noisy data}

To evaluate the stability of the reconstruction algorithm against random measurement errors, we add artificial noise to the clean data generated by \eqref{eq:Bdirekt}. On the real and imaginary parts of both magnetic field components, $B_x$ and $B_y$, we add Gaussian noise with zero mean value and with standard deviation determined as
\begin{align}
\sigma = \sigma_\text{ref}\bri{\brb{\Bvv_\text{meas}}} \label{eq:noise}
\end{align}
In \eqref{eq:noise}, $\sigma_\text{ref}$ is a reference parameter and $\bri{\brb{\Bvv_\text{meas}}}$ is the average over the measurement points of the magnitude of the magnetic field.

\begin{figure}[h]
\centering
\psfrag{R}{Re$\brc{I}$/A} \psfrag{I}{Im$\brc{I}$/A}
\includegraphics[scale=1]{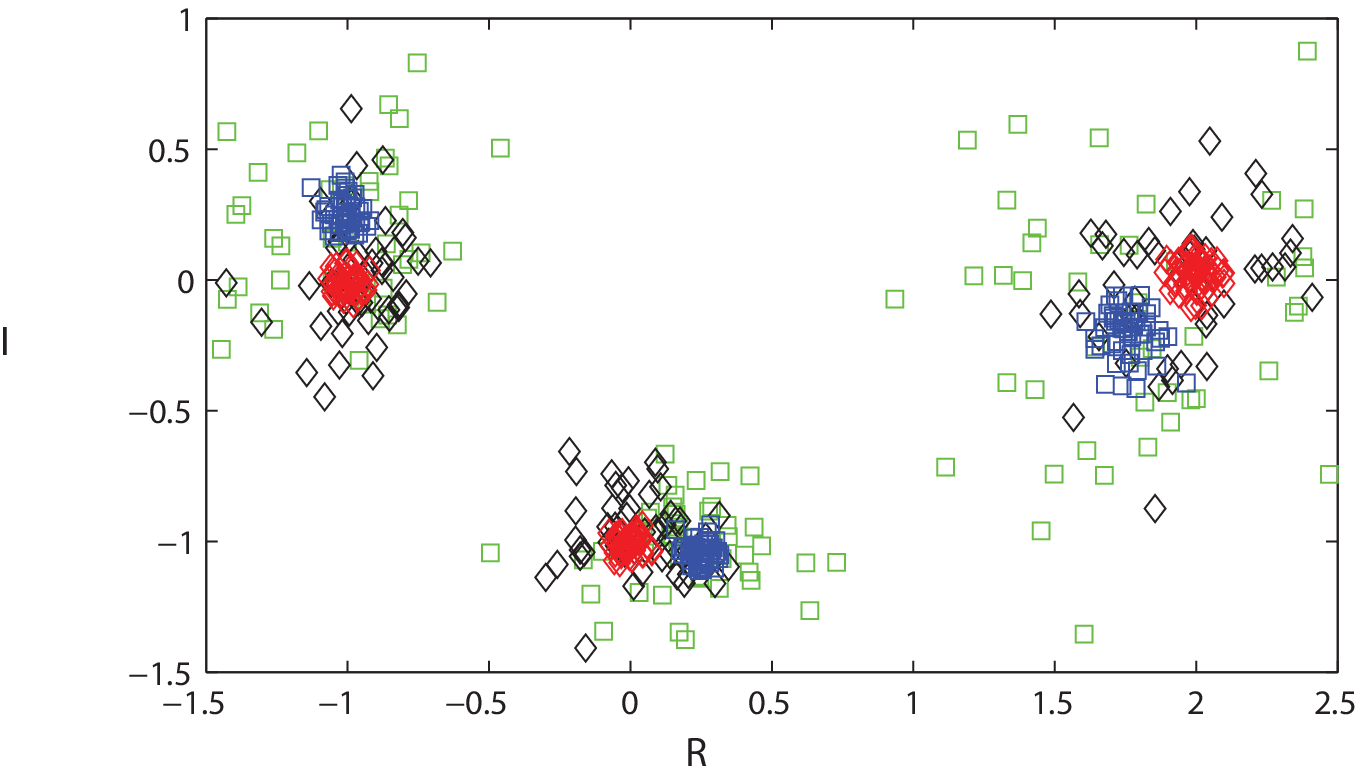}
\caption{Reconstructions of the phasor currents in the internal conductors using data contaminated with Gaussian noise. In each case we show 50 reconstructions. Green boxes: $N_\text{meas}=18$  and $\sigma_\text{ref}=0.05$. Blue boxes $N_\text{meas}=18$  and $\sigma_\text{ref}=0.01$. Black diamonds $N_\text{meas}=72$  and $\sigma_\text{ref}=0.05$. Red diamonds $N_\text{meas}=72$  and $\sigma_\text{ref}=0.01$. }\label{Fig:Irek}
\end{figure}

In the reconstruction examples, we consider two different levels of noise: $\sigma_\text{ref}=0.01$ and $\sigma_\text{ref}=0.05$, and for each level we test with $N_\text{meas}=18$ and $N_\text{meas}=72$ measurement points, respectively. To investigate the spread in the results we have for each case made 50 simulations.
\begin{figure}[h]
\centering \psfrag{x}{$x/R_\text{meas}$} \psfrag{y}{$y/R_\text{meas}$}
\includegraphics[scale=1]{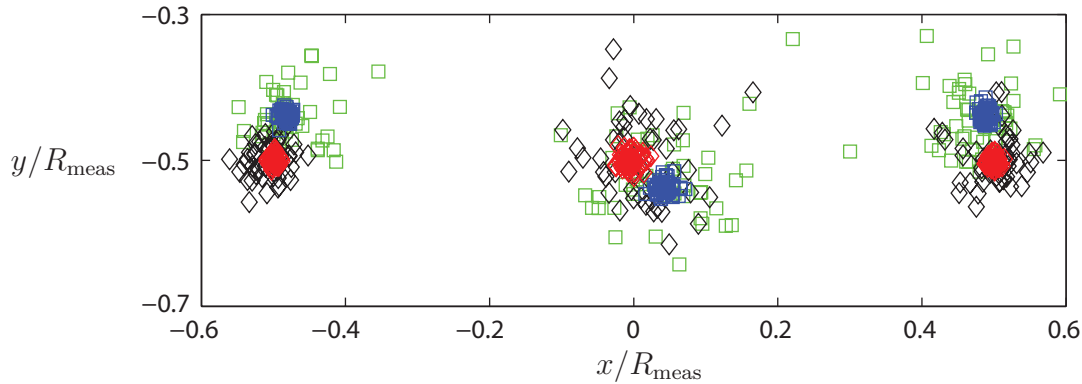}
\caption{Reconstructions of the positions of the internal conductors using data contaminated with Gaussian noise. In each case we show 50 reconstructions. Green boxes: $N_\text{meas}=18$  and $\sigma_\text{ref}=0.05$. Blue boxes $N_\text{meas}=18$  and $\sigma_\text{ref}=0.01$. Black diamonds $N_\text{meas}=72$  and $\sigma_\text{ref}=0.05$. Red diamonds $N_\text{meas}=72$  and $\sigma_\text{ref}=0.01$.   }\label{Fig:xyrek}
\end{figure}
The results for the conductor positions are presented in Figure \ref{Fig:xyrek} and the results for the complex phasor currents are presented in Figure \ref{Fig:Irek}. We see that increased noise levels increase the spread in the reconstructions. Increasing the number of measurement points has less impact on reducing the spread, but affects strongly the average positions, which tend to coincide with the results obtained when reconstructing from clean data; see Table \ref{tab:renadataxy}.

\section{Discussion}

\label{sekt:Discussion and conclusions}

Although the restriction to a line-current model has removed the non-uniqueness, the inverse problem is still illposed, in the sense that the algorithm is sensitive to both numerical errors and uncertainties in the measured data. Increasing the number of conductors, beyond three conductors considered in the numerical examples, the stability problem associated with the illposedness increases with an accelerating rate. Hence, if positions and currents are to be reconstructed simultaneously, a very large number of measurement points will be needed and the magnetic field sensors must have a very high precision; similar conclusions are drawn in \cite{CYGANSKI_ETAL:1984}.

The reconstruction algorithm has certain similarities with the algorithm used in \cite{CYGANSKI_ETAL:1984}, although their algorithm is based directly on point measurements. In our case, the construction of the algorithm from a Green identity yields that the integral in the right hand side of \eqref{eq:Greensksats8} has the property that the part of the magnetic field originating from currents outside the reconstruction domain gives no contribution; mathematically, our algorithm complete cancels the influence of disturbance sources.

As mentioned, in most electric power applications, the conductor positions are known or determinable by other means. Hence, despite difficulties with the full reconstruction problem, our algorithm can be applied to the reduced problem of reconstructing the currents only. In that case, the coefficients $f_m^n$ in \eqref{eq:Greensksats9} are evaluated from the known conductor positions, and the currents are determined directly from \eqref{eq:Iekv}. Thus, since the mathematical formulation cancels external disturbances, the measurement integral in \eqref{eq:Greensksats9} can be used as guideline for determining optimal sensors positions \cite{ZHANG_DIRIENZO:2009,DIRIENZO_ZHANG:2010,BRUZZONE_ETAL:2002} and for the processing of measurement data.

\section*{Acknowledgements}

This work is a part of the CIPOWER project within the EIT InnoEnergy programme.

\appendix

\section{On the non-uniqueness of the current density}

\label{app:tyst}

Here we give two examples of when the current density $J$ cannot be reconstructed uniquely using \eqref{eq:Greensksats6}.

\subsection{Surface currents}

Let $\mathcal{S}_0$ be a subsurface, completely interior to $\mathcal{S}$, bounded by the contour $\mathcal{C}_0$. We will show that on $\mathcal{C}_0$ one can always put a surface current density $\Kvv_0 = K_0\uz$ that exterior to $\mathcal{C}_0$ completely cancels the magnetic field from the current density $\Jvv_0=J_0\uz$ in $\mathcal{S}_0$. From \eqref{eq:divB}, the transversal magnetic field $\Bvv_0$ has the vector potential $\Avv_0=A_0\uz$. Hence, $\Bvv_0=\rot\Avv_0=\grad A_0\times\uz$, and using \eqref{eq:rotB} it follows that
\begin{align}
\lap A_0 = -\mu_0 J_0, \qquad \rvv\in \mathcal{S}_0
\label{eq:lapA0}
\end{align}
In consistence with no external field and absence of magnetic surface charge densities, we have on $\mathcal{C}_0$ that $\un_0\cdot\Bvv_0=\un_0\cdot\br{\grad A_0\times\uz} = \br{\uz\times\un_0}\cdot\grad A_0 = \uv{t}_0\cdot \grad A_0 = 0$. Hence,
\begin{align}
A_0 = a_0 = \text{constant}, \qquad \rvv\in \mathcal{C}_0
\label{eq:BC:A0}
\end{align}
\eqref{eq:lapA0} and \eqref{eq:BC:A0} are recognized as the interior Dirichlet problem for Poisson's equation. Hence, $A_0$ is uniquely determined in $\mathcal{S}_0$ expect for the undeterminable additive constant $a_0$ in \eqref{eq:BC:A0}. Since $a_0$ does not affect the magnetic field, we can set $a_0=0$. With the magnetic field determined uniquely, the required surface current density on $\mathcal{C}_0$, to achieve zero external field, follows from the tangential magnetic boundary condition:
\begin{align}
\Kvv_0 = \frac{1}{\mu_0}\un_0\times\br{\vc{0}-\Bvv_0} = -\frac{1}{\mu_0}\un_0\times\Bvv_0 = -\frac{1}{\mu_0}\un_0\times\br{\grad A_0\times\uz} = \frac{\uz}{\mu_0}\br{\un_0\cdot\grad A_0}
\end{align}
or
\begin{align}
K_0 = \frac{1}{\mu_0}\un_0\cdot\grad A_0
\label{eq:K0}
\end{align}
Hence, for an arbitrary choice of current density $J_0$ in $\mathcal{S}_0$ (which may also include surface and line currents) we always have a unique surface current density $K_0$ on $\mathcal{C}_0$ such that $J_0$ and $K_0$ produce no external field. This result is a two-dimensional analog of the magnetic signature reduction problem that has been discussed in \cite{NORGREN_HE_DEGAUSS:2000}.

Since $J_0$ and $K_0$ produce a zero field on $\mathcal{C}$, i.e. in the right hand side of \eqref{eq:Greensksats6}, their contribution in the left hand side of \eqref{eq:Greensksats6} must vanish, i.e.
\begin{align}
\int_{\mathcal{S}_0} J_0 \grad G \:\dS + \oint_{\mathcal{C}_0} K_0 \grad G \:\dl = \vc{0}
\label{eq:GS6lhs}
\end{align}
Expressed in index notation, \eqref{eq:lapA0} and \eqref{eq:K0}, together with the Gauss theorem, imply that \eqref{eq:GS6lhs} becomes proportional to
\begin{align}
 & -\int_{\mathcal{S}_0} \pden{G}{x_i}\pden{}{x_j}\pden{A_0}{x_j} \:\dS + \oint_{\mathcal{C}_0} \pden{G}{x_i}\pden{A_0}{x_j}n_j \:\dl \nonumber \\
= & -\int_{\mathcal{S}_0} \pden{G}{x_i}\pden{}{x_j}\pden{A_0}{x_j} \:\dS + \int_{\mathcal{S}_0} \pden{}{x_j}\br{\pden{G}{x_i}\pden{A_0}{x_j}} \dS = \int_{\mathcal{S}_0} \pden{A_0}{x_j}\pden{}{x_j}\pden{G}{x_i} \:\dS
\end{align}
Integration by parts, with $A_0=0$ on $\mathcal{C}_0$ and using \eqref{eq:lapG}, yields
\begin{align}
\int_{\mathcal{S}_0} \pden{A_0}{x_j}\pden{}{x_j}\pden{G}{x_i} \:\dS = - \int_{\mathcal{S}_0} A_0 \pden{}{x_j}\pden{}{x_j}\pden{G}{x_i} \:\dS = - \int_{\mathcal{S}_0} A_0 \pden{}{x_i}\lap G \:\dS = 0
\end{align}
which verifies \eqref{eq:GS6lhs}.

\subsection{Volume currents}

If we have a coaxial current density $J_0\brn{\rho}$, with support within the distance $a_0$ from a symmetry axis at $\rho=0$, that transports zero total current, i.e. fulfils the condition
\begin{align}
\int_0^{a_0}J_0\brn{\rho}\rho\:\dd\rho = 0
\label{eq:J0koax}
\end{align}
the magnetic field $\Bvv_0=\vc{0}$ when $\rho > a_0$. Hence, a superposition of such current densities, having parallel symmetry axes, produces no magnetic field outside its overall region of support. As a consistency check, we verify that a silent current density subject to \eqref{eq:J0koax} yields a zero contribution in the left hand side of \eqref{eq:Greensksats6}. Obeying the Laplace equation, $G$ has in the region $\rho < a_0$ the following general solution, in the polar coordinates $\brc{\rho, \vfi}$:
\begin{align}
G\brn{\rho,\vfi} = \sum_{m\in\mathbb{Z}} g_m \rho^{\brb{m}}\e^{\j m \vfi}
\end{align}
where $g_m$ are coefficients determined by the specific choice of $G$. Hence,
\begin{align}
\grad G = \uro\pdro{G}+\frac{\uvf}{\rho}\pdvf{G} = \sum_{m\in\mathbb{Z}} g_m\br{\brb{m}\uro+\j m \uvf}\rho^{\brb{m}-1}\e^{\j m \vfi}
\end{align}
Consequently, we must verify that
\begin{align}
\int_\mathcal{S}   J_0 \grad G \dS = \sum_{m\in\mathbb{Z}} g_m\int_0^{a_0} J_0\brn{\rho}\rho^{\brb{m}-1}\rho\:\dd\rho\int_0^{2\pi} \br{\brb{m}\uro+\j m \uvf}\e^{\j m \vfi}\dd\vfi   = 0
\label{eq:GJ0lhs}
\end{align}
Expressed in the fixed cartesian basis $\brc{\ux,\uy}$, the unit vectors $\brc{\uro,\uvf}$ are functions of $\e^{\pm \j\vfi}$. Hence, for $m\neq \pm 1$ the angular integral vanishes, while for $m=\pm 1$ the condition \eqref{eq:J0koax} applies to the radial integral, which verifies \eqref{eq:GJ0lhs}.

\bibliographystyle{IEEEtran}

\bibliography{IEEEabrv,Noncontact_ref}

\end{document}